# Comment on Lubineau et al. (2023) 'Does word flickering improve reading? Negative evidence from four experiments using low and high frequencies'

Michael Kodochian, CEO Abeye (Lexilens glasses), 27 rue Buffon, 21200 Beaune, France

## 1	Introduction

In 2023, Lubineau et al. published an article [1] detailing several experiments carried out with dyslexic readers. These authors attempted to measure the change in reading performance under different reading conditions using flickering devices. Beyond the low-frequency systems which have nevertheless shown their interest for some cases, we restrict here our response to the high-frequency systems, i.e. electronically controlled glasses (Lexilens) and lamps, designed and built upon recent work by Le Floch and Ropars [2]. Lubineau et al. found no significant change in reading performance at either low or high frequencies and concluded that these devices provide no or minor benefits. Unfortunately, experimental misunderstandings and some methodological issues invalidate namely the main conclusion concerning the high frequency systems.

## 2	Experimental misunderstanding

As already noted, we will not discuss here the low-frequency experiments 1 and 2 [1] which used different mechanisms from those used in our high-frequency systems as Lubineau et al. also wrote: "*Note that these results* [concerning experiments 1 and 2] *were obtained at low frequencies, quite far from the frequencies mentioned by Le Floch and Ropars, which were higher than 70 Hz"*. So, this comment only refers to experiments 3 and 4 which concern our devices.

First, in order to benchmark their reading performances, the article tries to compare the glasses and the lamps, in full daylight, "*to see if one of the two devices is more effective than the other*". As Lubineau et al. write "*all tests were therefore ran on paper, in a sufficiently bright room that did not require artificial lighting*". Unfortunately, the use of glasses is perfectly acceptable in the presence of standard daylight of 5000 lux, but by contrast with pulsed lamps as with all types of lamps, to be effective they must be used in darkness of less than about 50 lux. Furthermore, when using the pulsed lamps tested by Lubineau et al., even with a true continuous/pulsed ON/OFF system in a standard daylight, the role of the Hebbian mechanisms in the pulsed regime is inhibited by the additional daylight.

Second, the Lexilens glasses do not have a real ON/OFF. Indeed, there is an ON/OFF but it is only to save the battery. To test the two regimes, pulsed and continuous, the illuminances should be the same in both regimes to within 5%. However, in the false OFF position used by Lubineau et al., the glasses let in five times more light than in the ON position. Then, in the false ON/OFF comparison carried out by Lubineau et al., two parameters are changed at the same time, the regime (pulsed or continuous) but also the illuminance which is changed by 500%. Indeed, with glasses such as Lexilens which use liquid crystals between two crossed polarizers, the transmission in the OFF position is multiplied by a factor 5 compared to that in the ON position. Furthermore, it should be noted that in the Lubineau's test carried out in daylight *without any glasses*, in this case the illuminance is 1000% higher than with the glasses in the ON position, taking into account of a factor 2 due to the Malus' law through the input polarizer.

## 3   Methodological issues

- Experiment 3: impact of high-frequency flickering on a group of dyslexic students

The first test in Lubineau et al.'s experiment 3 is *naming letters aloud* (first graph in figure 4). Readers are instructed to read aloud single letters, that are well isolated one from each other. It is here expected that the readers' performance on this specific task is representative of their overall reading performance. Unfortunately, this is not the case. Indeed, it is well known that dyslexic readers perform similarly to normal readers when it comes to reading single letters [3–5]. In fact, when dyslexic children are asked to follow a target such as single letters without any cognitive task, the dyslexic and non-dyslexic groups cannot be distinguished.

Figure 4 also shows the change in performance in two other tasks, *single-word reading aloud* and *text reading aloud*. Word reading is a very complex task though that relies on several stores of information acquired over years of practice, i.e. the orthographic, the phonological and the semantic lexicons. The claim of the pulsed devices is to suppress the internal visual crowding due to extra mirror or duplicated images, i.e. to "clean" the visual input fed to downstream cognitive tasks. Unfortunately, it does not immediately correct the lexicons, which take a long time acquire. It therefore seems inappropriate to measure reading speed and error rates prior to catching up the same skills and lexicons that normal readers have acquired during their schooling.

- Experiment 4: single case study

In experiment 4, the authors focus on two subjects who have been using the Lexilens glasses for about two years. The results of Figure 6 in the objective and subjective tests could perhaps be significant with a pair of glasses with a true ON/OFF position. Unfortunately, here too, the false OFF position creates a situation where the illuminance in the continuous regime is 500% greater than that in the pulsed regime, thus invalidating all the conclusions of experiment 4.

These two subjects, FAP and CT have been surprised by the conclusions of Lubineau's article concerning them. Indeed, as they had told Lubineau and confirmed to us:

"I have been using these glasses for two years. I've noticed a clear difference between wearing them and not wearing them. When I read without my glasses, I have the impression that I have a series of black dots that interfere with my reading, and these dots are not there when I use my glasses. I don't get tired as quickly and I can read for 2 or 3 hours at a time with my glasses on, which is very difficult when I don't have them." (FAP's comment translated from French).

"I have been using my glasses since the fourth year of primary school. I am now (in 2024) in the 8$^{th}$ grade. At the time, reading was difficult because there was never enough space between letters. So I never read. When I put them on, the letters are bigger and more spaced apart. I am less tired when I use them. For the last 4 years, I have been reading a lot and I have been reading big novels." (CT's comment translated from French).

They do not understand why the tests carried out by Lubineau et al. fail to confirm the beneficial effect of these glasses, which they of course continue to use every day. It must be said that here too, the false ON/OFF was used, leading to unusable data.

## 4  Conclusion

Unfortunately, the standard tests carried out by Lubineau et al. contain experimental errors that lead to statistically insignificant results. Moreover, in order to carry out such tests, serious prior remediation is required to ensure that the child has acquired orthographic, phonological and semantic lexicons comparable to those of normal readers. The correct use of the pulsed systems, once optimised, essentially eliminates the noisy internal visual crowding due to the extra mirror or duplicate images. Tests to demonstrate the effectiveness of pulsed systems must be designed to observe the immediate disappearance of this internal visual crowding. For example, the recent eye tracker experiment [6] demonstrates the effectiveness of pulsed screens. In fact, the excess of eye fixations and total reading times observed in many laboratories [7–10] with dyslexic readers can be controlled by simply switching to the pulsed mode. High-frequency systems which use Hebbian mechanisms at the synapses of the primary cortex are necessary to suppress the internal visual crowding and are useful systems for helping dyslexic children and adults, especially at high frequencies above 60/70 Hz, as any annoying flicker is completely invisible to the human eye. Of course, before dyslexic children can read fluently, they still need to learn the correct lexicons, as their vocabulary needs to be strengthened.

Moreover, Lubineau et al. declare no competing interests. However, the main author (Lubineau) is an employee at CERENE, a network of private schools in France, specialised in teaching to children with learning disorders, founded and directed by H. Glasel, co-author of the article. The Lexilens glasses could indeed be considered a competitor to CERENE business. In the interest of the dyslexic children, it might have been preferable to try to include these pulsed systems in the panoply of the techniques they use. In fact, many people in different countries around the world already use the Lexilens glasses every day with satisfaction.

**Data accessibility**. This article has no additional data.

**Declaration of AI use**. We have not used AI-assisted technologies in creating this article.

**Conflict of interest**. The author is the CEO of the Abeye corporation which developed the Lexilens glasses.

**Funding**. This work received no funding.

1. Lubineau M, Watkins CP, Glasel H, Dehaene S. 2023 Does word flickering improve reading? Negative evidence from four experiments using low and high frequencies. *Proc. R. Soc. B Biol. Sci.* **290**, 20231665. (doi:10.1098/rspb.2023.1665)

2. Le Floch A, Ropars G. 2017 Left-right asymmetry of the Maxwell spot centroids in adults without and with dyslexia. *Proc. R. Soc. B* **284**, 20171380. (doi:10.1098/rspb.2017.1380)

3. Adler-Grinberg D, Stark L. 1978 Eye Movements, Scanpaths, and Dyslexia. *Optom. Vis. Sci.* **55**, 557.

4. Prado C, Dubois M, Valdois S. 2007 The eye movements of dyslexic children during reading and visual search: impact of the visual attention span. *Vision Res.* **47**, 2521–2530. (doi:10.1016/j.visres.2007.06.001)

5. Bucci MP, Nassibi N, Gerard C-L, Bui-Quoc E, Seassau M. 2012 Immaturity of the oculomotor saccade and vergence interaction in dyslexic children: evidence from a reading and visual search study. *PloS One* **7**, e33458. (doi:10.1371/journal.pone.0033458)

6. Le Floch A, Ropars G. 2023 Hebbian Control of Fixations in a Dyslexic Reader: A Case Report. *Brain Sci.* **13**, 1478. (doi:10.3390/brainsci13101478)

7. Rello L, Ballesteros M. 2015 Detecting readers with dyslexia using machine learning with eye tracking measures. In *Proceedings of the 12th International Web for All Conference*, pp. 1–8. New York, NY, USA: Association for Computing Machinery. (doi:10.1145/2745555.2746644)

8. Benfatto MN, Seimyr GÖ, Ygge J, Pansell T, Rydberg A, Jacobson C. 2016 Screening for Dyslexia Using Eye Tracking during Reading. *PLOS ONE* **11**, e0165508. (doi:10.1371/journal.pone.0165508)

9. Asvestopoulou T, Manousaki V, Psistakis A, Smyrnakis I, Andreadakis V, Aslanides IM, Papadopouli M. 2019 DysLexML: Screening Tool for Dyslexia Using Machine Learning. (doi:10.48550/arXiv.1903.06274)

10. Raatikainen P, Hautala J, Loberg O, Kärkkäinen T, Leppänen P, Nieminen P. 2021 Detection of developmental dyslexia with machine learning using eye movement data. *Array* **12**, 100087. (doi:10.1016/j.array.2021.100087)